# Novel Cyano-Bridged 4f-3d Coordination Polymers with a Unique 2D Topological Architecture and Unusual Magnetic Behaviors**

**Bao-Qing Ma, Song Gao,* Gang Su, and Guang-Xian Xu**

[*]  Prof. Dr. S. Gao, Dr. B.-Q. Ma, Prof. Dr. G.-X. Xu

State Key Laboratory of Rare Earth Materials Chemistry and Applications, PKU-HKU Joint

Laboratory on Rare Earth Materials and Bioinorganic Chemistry

Peking University

Beijing 100871 (P. R. China)

Fax: (+86) 10-6275-1708

E-mail: gaosong@chemms.chem.pku.edu.cn

Prof. Dr. G. Su

Department of Physics, Graduate School at Beijing, University of Science and Technology

of China, CAS, P.O. Box 3908, Beijing 100039 (P. R. China)

[**]This work was supported by the State Key Project of Fundamental Research (G1998061305), the National Natural Science Foundation of China (29771001 and 29831010) and the Excellent Young Teachers Fund of MOE, P. R. C..

Cyano-bridged bimetallic hybrid Prussian Blue 1-3D coordination polymers based on $[M(CN)_6]^{3-}$ (M = Fe, Cr, Mn) have attracted great attention due to their rich and interesting structures and magnetic behaviors.[1] These studies were mainly focused on transition metals. In principle, it would be likely to enhance the coercive fields by the introduction of paramagnetic lanthanide ions because the latter possesses rather large and anisotropic magnetic moments. However, the magnetism, yet rarely investigated in literature, for known cyano-bridged lanthanide and transition metal complexes such as ion pairs, dinuclear, trinuclear, tetranuclear, and one-dimensional chains,[2] etc., seems not to be exciting, since the couplings between lanthanide and transition metals are considerably weak, which arises from the effective shielding of 4f electrons by the outer shell electrons. On the other hand, it was noted that the 3D $SmFe(CN)_6 \cdot 4H_2O$, with strong anisotropic coercive fields, exhibits a long range ferrimagnetic ordering below 3.5 K, and $TbCr(CN)_6 \cdot 4H_2O$ has the highest $T_C$ (11.7K) in 4f-3d molecule-based magnets up to now[3]. This implies that to increase dimensionality may enhance and improve bulk magnetic properties. Our strategy for the rational synthesis of high-dimensional network is to make suitable combination of cyanides and bridging ligands. In the present work, 2,2'-bipyrimidine (bpym) is selected, because it is more capable of transmitting magnetic interactions[4] and bis(chelating) coordination modes which facilitates the connection between lanthanide ions than 4,4'-bipyrazine and pyrazine. Unexpectedly, two novel coordination polymers $[NdM(bpym)(H_2O)_4(CN)_6] \cdot 3H_2O$ (M = Fe **1**, Co **2**;) have been obtained, which have a unique 2D topological architecture, and exhibit some unusual magnetic behaviors.

Compounds **1** or **2** were obtained by slow diffusion of $K_3Fe(CN)_6$ or $K_3Co(CN)_6$ and bpym with $Nd(NO_3)_3$ in 1:1:1 molar ratio in 30mL aqueous solution. X-ray diffraction analysis reveals that they are isomorphous.[5] An ORTEP drawing of **1** is shown in Figure 1. It consists of a two-dimensional net with alternating fused rows of rhombus $\{Fe_2Nd_2(CN)_4\}$ rings and six-sided $\{Fe_4Nd_4(CN)_8\}$ rings. Each $Nd^{3+}$ is eight coordinated by two N atoms from one chelating bpym molecule (the mean separation of Nd-N = 2.7084 Å), four O atoms from water molecules (the mean separation of Nd-O = 2.4706 Å) and three N atoms from three bridging $CN^-$ groups (the mean separation of Nd-N = 2.5825 Å), yielding a mono-capped square anti-prism. The top and bottom planes are defined by O1, O2, N2, N10a and O3, O4, N1, N5, respectively, and N8 occupies the cap position. The bpym ligand coordinates to Nd ions in a chelating fashion, and the remaining two N atoms form the hydrogen bondings with the coordination water molecules of neighboring $Nd^{3+}$ ion (O3···N3a = 2.929 Å, O3-H5···N3a = 155.80°; O2···N4a = 2.850 Å, O2-H3···N4a = 160.08°. symmetry code: a x, y+1, z). Although the chelating mode of bpym and 2,2'-bipyridine have similar steric effects, they form different structures with $K_3Fe(CN)_6$ and $Nd(NO_3)_3$. For the latter, an ion pair compound $[FeNd_2(CN)_6(2,2'-bipy)_4(H_2O)_8][Fe(CN)_6].8H_2O$ is given.[2f] The difference may be tentatively attributed to the different affinity of both ligands.

There exist two crystallographically unequal Fe atoms in an asymmetric unit. They are all located in an inversion center. Fe2 employs four $CN^-$ groups in the same plane to connect the $Nd^{3+}$ ions, giving rise to a double strange chain as found in $[\{Cu(dien)_2Cr(CN)_6\}_n]$ $[Cu(dien)(H_2O)Cr(CN)_6]_n \cdot 4nH_2O$ (dien = diethylenetriamine)[6]. Thus, a 12-membered rhombus subunit $\{Fe_2Nd_2(CN)_6\}$ motif is generated with Nd1···Fe2 and Nd1···Fe2a separations being 5.616 and 5.606 Å, respectively, similar to that observed in $[Fe_2(CN)_4(phen)_4Yb_2Cl_6(H_2O)_2]\cdot2H_2O\cdot2CH_3OH$.[2l] Fe1 uses two *trans* CN groups to link these neighboring chains with Nd1···Fe1 separation being 5.381 Å, leading to an unusual 2D sheet as shown in Figure 2a. The shortest interlayer Nd···Fe, Nd···Nd and Fe···Fe distances are 9.390, 8.897 and 9.197 Å, respectively. The Fe-C distances are in the range of 1.9171-1.9450 Å and Fe-C≡N bond angles are nearly linear. The bridging cyanide ligands link the $Nd^{3+}$ ions in a bent fashion with the Nd-N≡C bond angles ranging from 158.52(9)° to 166.11(11)°. The lattice water molecules reside between layers and link the 2D sheets into a 3D network through rich hydrogen bondings.

The previously reported cyanide-bridged 4f-3d complexes just display a 0D or 1D structural character. It turns out that the compounds **1** and **2** reported in the present work serve as the first cyanide-bridged two-dimensional compound containing both lanthanide and transition metal ions. More interestingly, they supply a novel 2D topological architecture type, pronouncedly different from the known 2D cyanide-bridged square,[7] honeycomb[8], brick wall[9], partial cubane structures,[10] and other 2D coordination polymers (grid, kagome, herring bone, triangular).[11]

The temperature dependences of the $\chi_M T$ for crystalline samples **1** and **2** confined in parafilm, measured at 10 kOe and 500 Oe fields, are shown in Figure 3 and in the inset of Fig 3, respectively. The observed $\chi_M T$ values for **1** and **2** at 300 K are 1.87 and 1.49 $cm^3$ $mol^{-1}$ K, slightly smaller than the calculated values 2.00 and 1.64 $cm^3$ $mol^{-1}$ K for non-interacting free ions per NdFe and NdCo unit ($Co^{3+}$ is diamagnetic in the present strong field case), respectively. Upon cooling, the $\chi_M T$ of **2** decreases monotonously from 300 to 2 K, while the $\chi_M T$ of **1** is roughly parallel with that of **2** above 80 K. The difference is approximately constant and equal to 0.38 $cm^3$ $mol^{-1}$ K, which represents the values of an isolated iron(III) ion. These observations suggest that the reduction of $\chi_M T$ with lowering temperature must be considered as an intrinsic characteristic of Nd ion, which may be essentially attributed to the depopulation of its Stark levels.[12a] When further cooling, the $\chi_M T$ of **1** reaches a minimum at ca. 10K, then goes up fast. To exclude the contribution of spin-orbital coupling of $Nd^{3+}$, the $\chi_M T$ of **2** was subtracted from that of **1.** It is found that the difference (solid line shown in the inset of Figure 3) increases monotonously with cooling, an indicative of ferromagnetic interactions between $Nd^{3+}$ and $Fe^{3+}$ ions. No long range ordering was observed down to 2 K based on measurements of the field dependent magnetization and the zero-field AC susceptibility, however. The field dependence of magnetization *M* for **1** and **2** was measured at 1.8 K. The magnetization values at 50 kOe are 2.06 and 1.05 Nβ, respectively, close to saturation. For $Nd^{3+}$ in **2**, the crystalline field resolves the 10-fold degenerate $^4I_{9/2}$ ground states into five doublets, and only one of which is populated at temperature 20K or below.[12b] If the saturation magnetization value of **2** is taken as 1.09 Nβ (60kOe), the effective spin in the ground state for one $Nd^{3+}$ ion is ½, and g' is 2.18, consistent with the above argument. The difference *M* for **1** and **2** yields 1.01 Nβ for $Fe^{3+}$, in fair accord with its theoretical value 1 Nβ.

For simplicity, compound **2** containing only one type of paramagnetic $Nd^{3+}$ ion was investigated in some detail. The zero field AC susceptibility was measured in 3-8 K range at different frequencies (111-1111Hz), and shown in the upper panel of Figure 4. Normal paramagnetic behavior of the in-phase component $\chi_M$' and negligible small signal of the out-of-phase component $\chi_M$", indicate the absence of long range ordering and any spin glass-like behavior. However, under an intermediate DC bias field (herein 2 kOe), two peaks ($T_P$) appear for both $\chi_M$' and $\chi_M$", respectively, as shown in the lower panel of Figure 4. Moreover, $\chi_M$' and $\chi_M$" are strongly frequency dependent with a considerably larger $\phi$ value 0.37 ($\phi = \Delta T_P/(T_P(\log f))$) than that (<0.1) for normal spin glass,[13] suggesting the unusual glassy-like behavior, which may be attributed to the geometrical frustration of $Nd^{3+}$ ions as shown in Figure 2b. The 2D connection of $Nd^{3+}$ can be viewed as two penetrated sets of honeycombs (red and black) with very similar linear N-C-Co-C-N five-atom bridges close to 11Å. If just one set of honeycomb such as black one is considered, the antiferromagnetic (AF) coupling between neighboring $Nd^{3+}$ ions should not give rise a frustration. However, since the pathways between the two sets of honeycombs are also coupled through N-C-Co-C-N five-atom bridges, the magnetic interaction between the neighboring spins on different honeycombs might be comparable with that in each honeycomb, leading to that the magnetic moments located in $Nd^{3+}$ will be frustrated.

Experimentally, at zero field no frustration was observed for **2** owing to rather weak AF interactions between $Nd^{3+}$ ions, whereas the introduction of a DC field enhances the AF correlation, which gives rise to the unusual field-dependent magnetic relaxation. The further study is in progress.

*Experimental section*

bpym was purchased from Aldrich and used without further purification. $K_3Co(CN)_6$ was prepared by the method reported in literature.[14] Compound **1** was prepared by mixing $K_3Fe(CN)_6$ (0.5mmol) and bpym (1mmol) in 20mL aqueous solution, followed by the slow addition of 10 mL aqueous solution of $Nd(NO_3)_3$(1mmol) without any stirring. The resulting solution was allowed to slowly diffuse and evaporate at room temperature in the dark. After two weeks, bright yellow single crystals were obtained, which were collected by filtration, washed with small amounts of water and ethanol, and dried in air. Calcd. for $C_{14}H_{20}FeNdN_{10}O_7$: C, 26.23; H, 3.12; N, 21.86. Found: C, 26.86; H, 2.78; N, 22.26%. Compound **2** was synthesized following the same procedure as **1**. Only the $K_3Fe(CN)_6$ was replaced by $K_3Co(CN)_6$. Calcd. for $C_{14}H_{20}CoNdN_{10}O_7$: C, 26.09; H, 3.11; N, 21.77. Found: C, 26.43; H, 2.83; N, 22.16%.

**Keywords:** cyanide • magnetic properties • crystal structure • lanthanide complexes • layer compounds

---

based on $F^2$ using SHELXL 97 program. Crystallographic data (excluding structure factors) for structures **1** and **2** have been deposited with the Cambridge Crystallographic Data Centre as supplementary publication no. CCDC-147180 and 147181, respectively. Copies of the data can be obtained free of charge on application to CCDC, 12 Union Road, Cambridge CB21EZ, UK (Fax: (+44) 1223-336-033; E-mail: deposit@ccdc.cam.ac.uk).

# Figure captions

**Figure 1.** Local connections of [NdFe(bpym)(H$_2$O)$_4$(CN)$_6$]·3H$_2$O **1**. Selected bond lengths [Å] and angles [°]: Nd(1)-O(1) = 2.4235(10), Nd(1)-O(2) = 2.4524(11), Nd(1)-O(3) = 2.4739(11), Nd(1)-N(5) = 2.5105(10), Nd(1)-O(4) = 2.5326(10), Nd(1)-N(10a) = 2.5972(15), Nd(1)-N(8) = 2.6399(13), Nd(1)-N(1) = 2.6875(13), Nd(1)-N(2) = 2.7293(10) (Symmetry codes: a x+1, y, z).

**Figure 2. a** Layer structure of **1**, exhibiting a unique topological architecture. For clarity, only the cyanide groups are shown. **b** The interaction between two sets of honeycombs in 2D connection of Nd$^{3+}$ leading to spin frustration in compound **2**.

**Figure 3.** Temperature dependence of $\chi_M$T for **1** (O) and **2** (  ) at 10K Oe. Inset is the curves of $\chi_M$T versus T at 500 Oe.

**Figure 4.** Frequency dependence of Ac susceptibility at zero and 2K Oe DC bias field for compound **2**.

**Figure S1.** Ac susceptibility of compound **1** under zero-field at different frequencies.

**Figure S2.** Field dependence of the magnetization M at 1.8K for **1** (O) and **2** (  ).

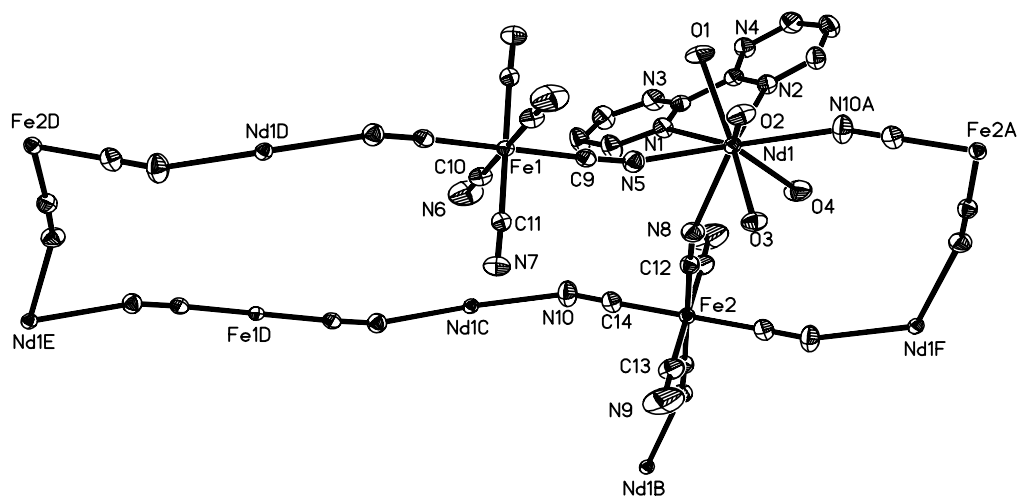

Figure 1

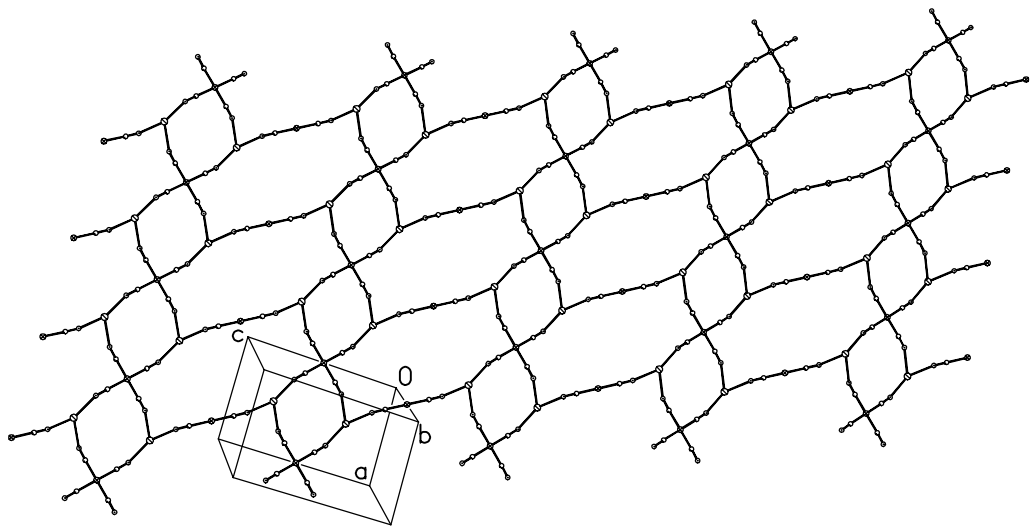

**(a)**

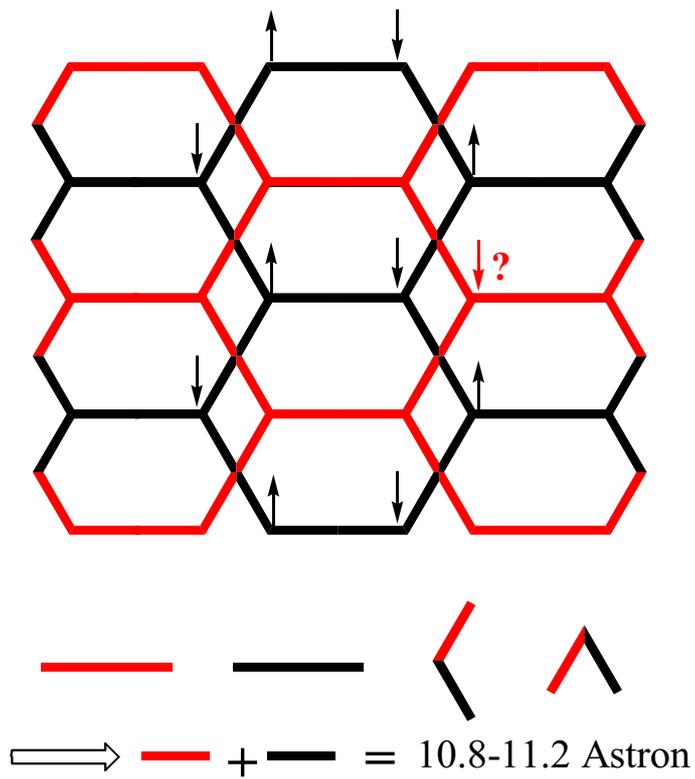

= 10.8-11.2 Astron

**(b)**

Figure 2

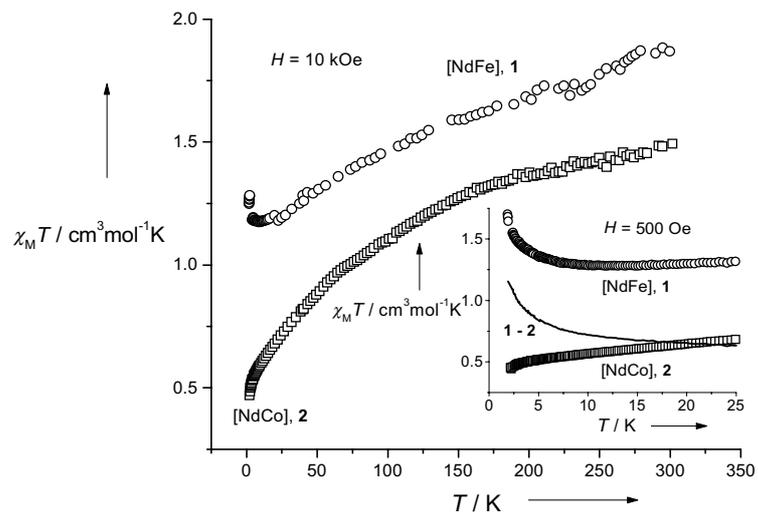

Figure 3

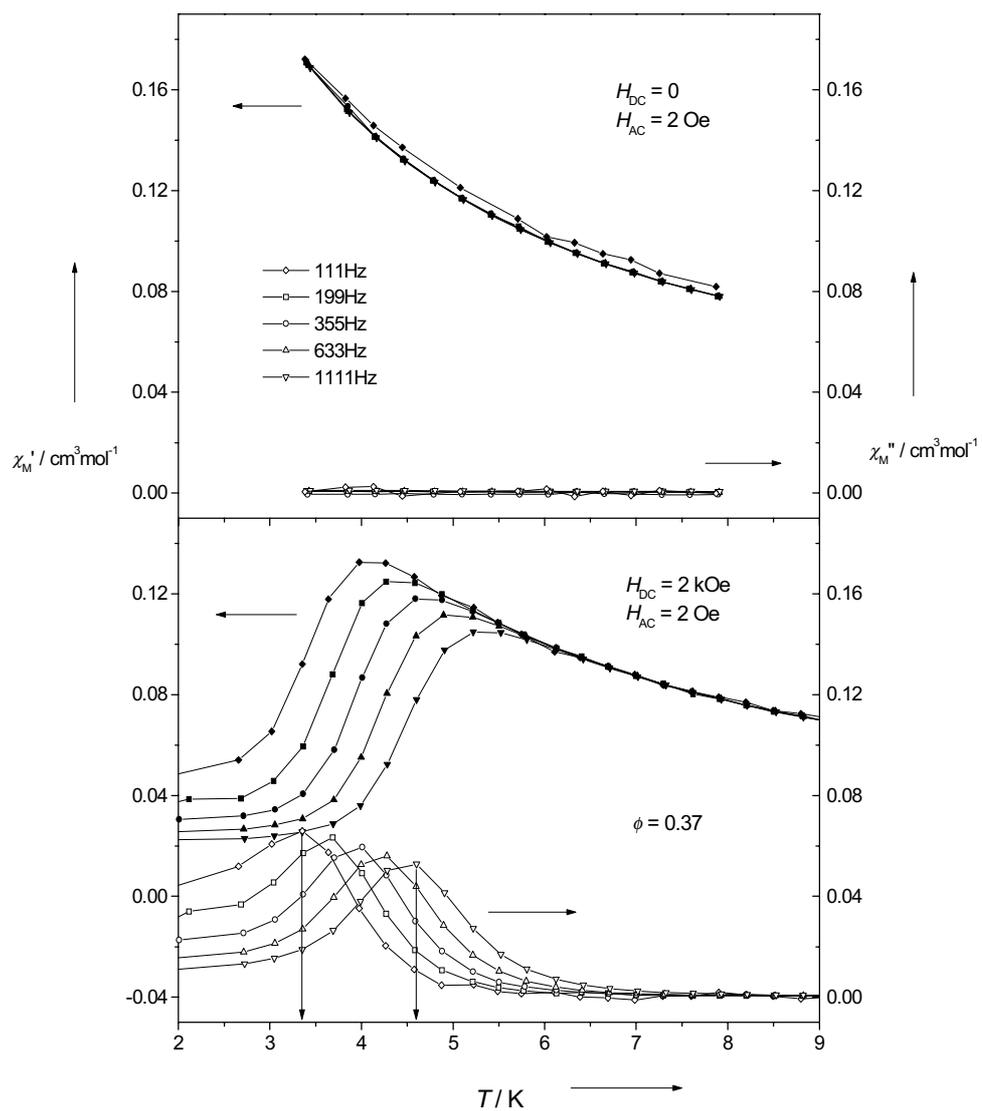

Figure 4.

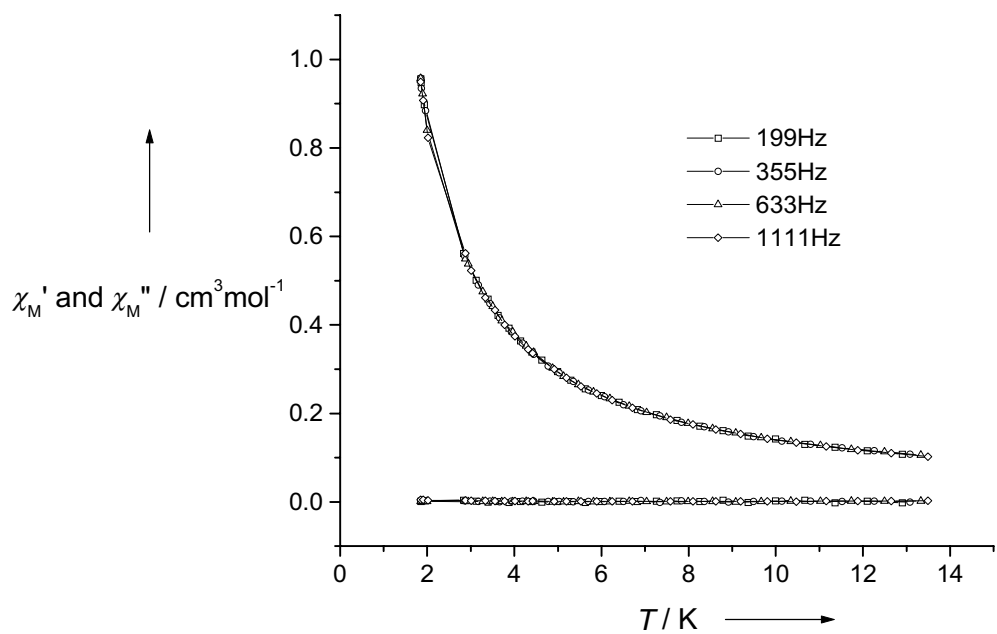

**Figure S1**

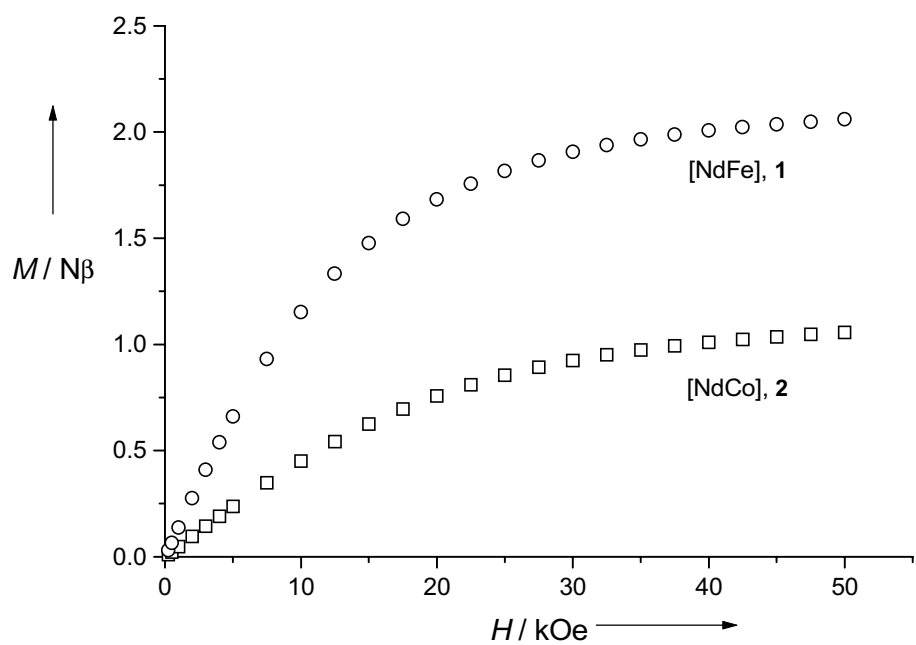

**Figure S2**

Novel Cyano-Bridged 4f-3d Coordination Polymers with a Unique 2D Topological Architecture and Unusual Magnetic Behaviors

Bao-Qing Ma, Song Gao, Gang Su and Guang-Xian Xu

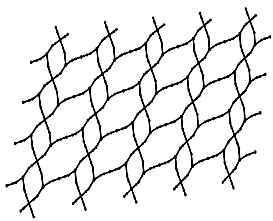

**Two novel compounds** [NdM(bpym)(H$_2$O)$_4$(CN)$_6$] 3H$_2$O (M = Fe **1**, Co **2**), serving as the first cyano-bridged two-dimensional 4f-3d complexes, exhibit a unique topological architecture type, and unusual magnetic behaviors.